# Graphene channels interfaced with an array of individual quantum dots

Xin Miao and Haim Grebel*

*Electronic Imaging Center and ECE Dept., New Jersey Institute of technology (NJIT), Newark, NJ 07102, USA.*

**Abstract** – Surface graphene guides were interfaced with an array of individual semiconductor quantum dots, whose position was commensurate with the optical guide modes. The surface guide served as a channel for a Field Effect Transistor (FET) while the dots were placed within the capacitor formed between the graphene channel and the gate electrode. We report on negative differential photo-related conductance under light and a diminishing fluorescence effect as a function of bias. We suggest that the quenched fluorescence may be hindered, to some degree, by incorporating the QD in a resonator, which is tuned to the emission wavelength.

Keywords: Surface optical guides; field effect transistors (FET); graphene channels; semiconductor quantum dots (QD);

*Email: grebel@njit.edu

Graphene [1], a mono, or a few layers of graphite, has attracted a vast interest recently [2,3]. Early on, field effect transistors (FET) demonstrated its unique electrical properties [4]. One may also expect unique electrical effects when the graphene (or graphene oxide, GO) is interfaced with semiconductor quantum dots (QDs) [5-8] or with dyes [9-12]. In those experiments, the fluorophores were placed on top of the graphene substrates, contrary to the present design. Graphene quenched the fluorescence and recent interpretations attributed it to a physical transfer of electrons from the fluorophores to the graphene [7,9], similarly to donor doping in semiconductors. Somewhat in support of that notion was given in [8]; the fluorescence quenching hindered as the distance between QD and GO increased. Similar to SWCNT, the mobility of a graphene coated with an optically sensitive film ought to depend on the mobility of carriers in the film as well [13]. We note that: (a) the photo-induced transport as a function of light intensity involves the entire graphene channel and (b) the channel characteristics nearby the QD is more local and may directly affect the fluorescence process [14]. A different point of view was given in [15]; the energy transfer between the QD and graphene is attributed to FRET (frequency resonance energy transfer which is enabled through screening by the graphene). The problem is that near the Dirac point such screening is linearly diminishing [16] and the screening, if exists, should be non-linear and depending on the amount of charge placed within a small distance away from the graphene [17, 18]. While not directly related to the quenching mechanism(s), we set here to investigate the effect of bias on the photo-conductivity and fluorescence of individual QD when placed within the gate-channel capacitor.

Electrical properties of graphene on periodic and porous substrates, such as anodized aluminum oxide were studied in the past [19-20]. It was found that the periodic holes array may accentuate the Raman spectra of the graphene lines and led to the realization of the first visible surface plasmon laser [21-23]; there, one takes advantage of simultaneous resonating plasmon/polariton modes at both the pump and at the emission frequencies. Here we go one step further and focus on the electro-optical and photoluminescence as a function of the device bias; by suppressing the coupling between the pump laser radiation and its related propagating surface modes we concentrate on only the emission radiation. Additionally, since the graphene is partially suspended over the substrate pores, the characteristic parameter $\alpha=e^2/(\varepsilon\hbar v_F)>1$ with $\varepsilon$, the dielectric constant of the vacuum [17]. Finally, the absorption of graphene (~2.3% per layer) is comparable to the absorption of monolayer of CdSe/ZnS QD (the linear absorption coefficient of QD is $A\sim10^5$ /cm and a typical dot diameter is D=3 nm. If we assume that the absorption behaves as [1-exp(A*L)]~A*L, then the absorption of a QD monolayer is ~3%).

At visible and near-IR wavelengths, graphene acts as a lossy dielectric [24]. Since the graphene is atomically thin, we were able to realize a surface guide, sandwiched between two lower dielectric media: air/polymer on the top and silica/alumina at its bottom. At the same time, the graphene's conductivity may be tuned by biasing. This enabled us studying the effect of varying conductivity on the optically induced current and on the related QD photoluminescence. The array of pores in the anodized aluminum oxide layer provided us with yet another advantage. Surface modes decay exponentially away from the thin guide and, hence are concentrated at the guide surface (and toward the QDs). The periodic pattern of pores enabled coupling between the free space radiation and the propagating surface modes. If properly designed, the array of pores may facilitate standing surface modes for a strong coupling between electromagnetic radiation and QDs [25].

The schematic of the FET and an SEM picture of the porous substrate are shown in Fig. 1. 20 nm of $SiO_2$ (or in some cases, 150 nm) of oxide was deposited on a <100> p-type 1-10 Ohms.cm Si wafer; the Si served as a back gate electrode. For the anodization, a 1-micron Al film was deposited on top of the $SiO_2$ layer; the Al was later anodized completely per previous recipe [26] – its final thickness was estimated as ~50 nm. Anodization of the Al resulted in a hole-array with a pitch of ca 100 nm and a hole-diameter of less than 30 nm. The hexagonal hole-array was polycrystalline with a typical domain size of ~10 microns. The CdSe/ZnS QD either with peak luminescence at 590 nm, or at 670 nm were suspended in toluene and drop-casted into the anodized porous substrate. The QDs were coated with octadecylamine to prevent agglomeration while in suspension. Mostly one QD occupied a filled AAO nano-hole (Fig. 1b). Excess dots lying on the substrate surface were washed away. The graphene was produced by chemical vapor deposition technique (CVD) on copper foil and was transfer onto the QD embedded substrate by use of 250 nm poly(methyl methacrylate), PMMA film [27]. We retained the PMMA film as a protective upper coating and as a dielectric layer. The deposition method yielded no more than a 2-layer film, as determined by Raman spectroscopy. Raman spectroscopy of the QD interfaced graphene also revealed a large graphene defect line, situated at 1340 $cm^{-1}$. This line is rather small for a free-standing graphene, or graphene films deposited on quartz. The linear $I_{ds}$-$V_{ds}$ curve may be explained by the large surface states at the area of contact making it ohmic. Luminescence data were obtained in confocal arrangement. A 30 mW, CW, 532nm Nd:YAG laser was focused to a 25 $\mu m^2$ spot. The sample was tilted and rotated to produce optimal coupling with the surface modes as in [22].

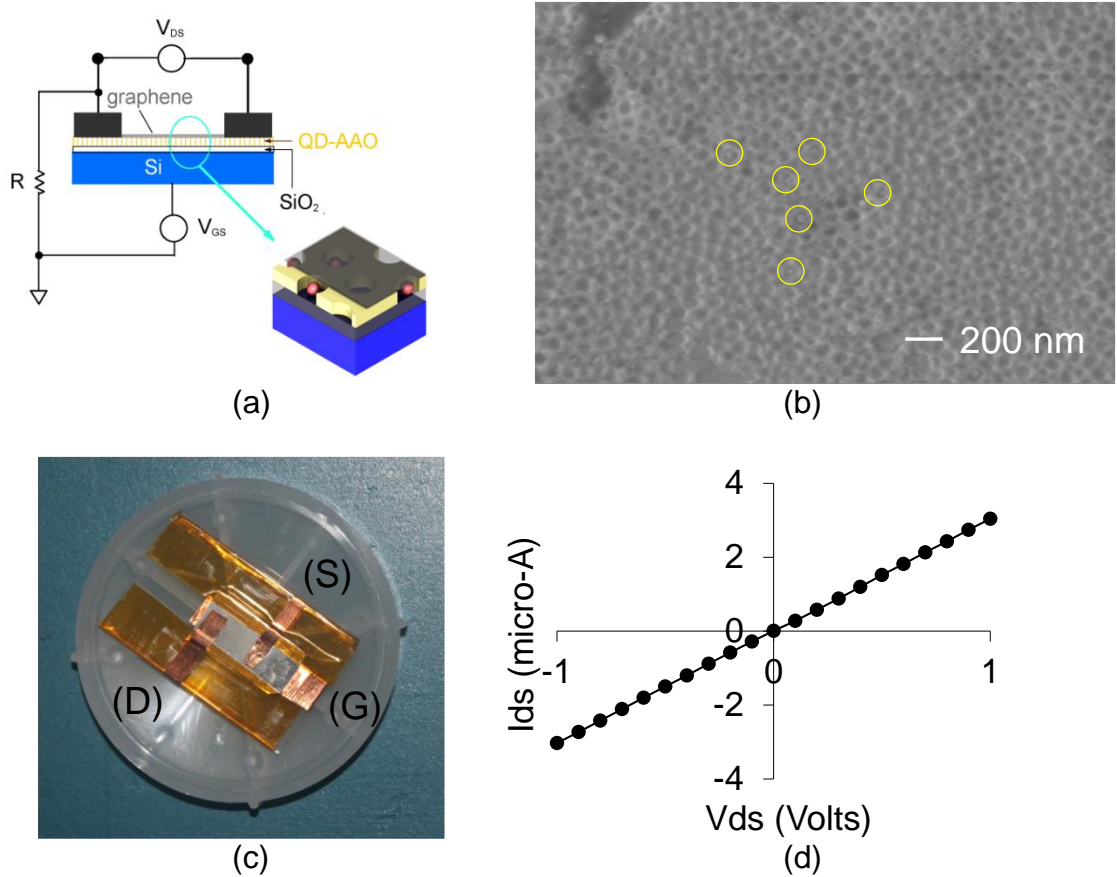

Fig. 1. (a) Schematics of the device configuration. (b) SEM picture of QD-filled AAO. The position of a dot is marked by a yellow circle. (c) 1 cm² channeled device with a transferred CVD grown graphene: (D), (S) and (G) are Drain, Source and Gate electrodes, respectively. The graphene was depositing the region between the D and S on top of the Cu electrodes. (d) A linear $I_{ds}$-$V_{ds}$ curve of the graphene channel attests to its ohmic contacts.

The channel became more conductive as a function of both $V_{gs}$ and $V_{ds}$ when uniformly illuminated by a 30 mW/cm², CW Nd:YAG laser at 532 nm (Fig. 2). This intensity is much smaller than had been used by either Ref. 7 or 9. The major dip in the illuminated curve at $V_{gs}$=−2.3 V can be identified as the position of the Dirac point, which has been shifted from $V_{gs}$=−1.3 V for the non-illuminated case. The second dip, at $V_{gs}$=−2.7 V is shared by the channel under dark conditions and hence can be attributed to the effect of $V_{gs}$ on the QD band structure itself. The third dip at $V_{gs}=-4.6$ V appears only for the illuminated curve (for all $V_{ds}$ values) and may be attributed to negative differential photo-conductance [28].

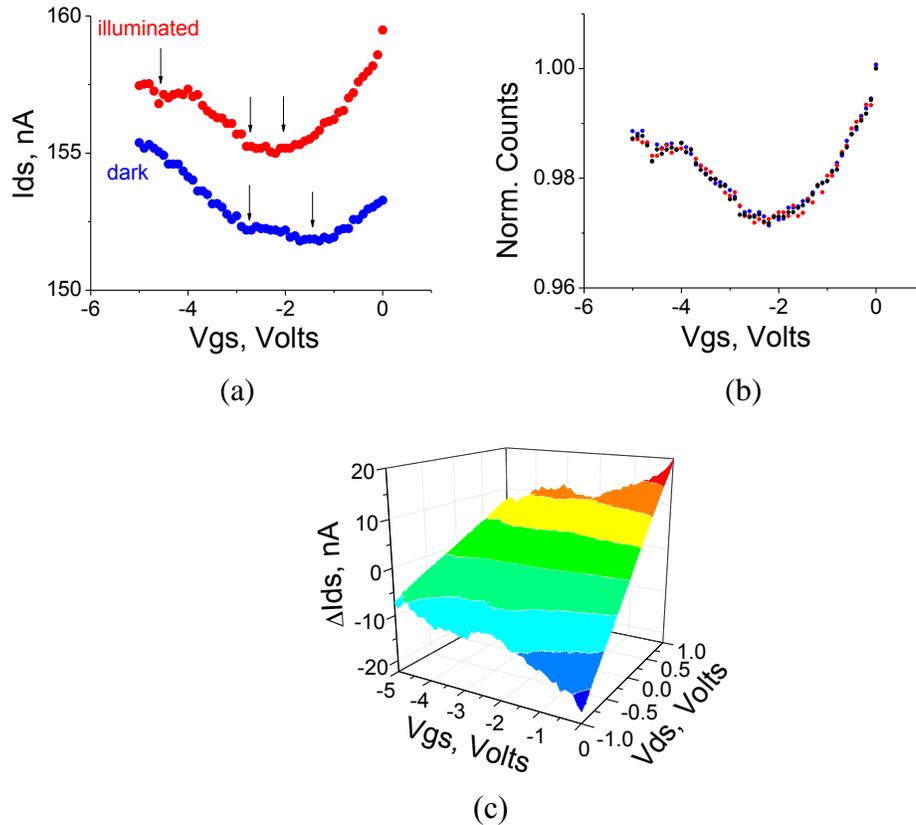

(a)

(b)

(c)

Fig. 2. Channel conductivity under dark and under uniform illumination by a 30 mW/cm$^2$ 532 CW laser. (a) Comparison between illuminated and non-illuminated $I_{ds}$-$V_{gs}$ curves. $V_{ds}=0.3$ V. The arrows point to the position of the various dips. (b) Normalized $I_{ds}$-$V_{gs}$ curves: illuminated sample at $V_{ds}=0.1$, 0.2 and 0.3 V. (c) The differential current (channel current difference between illuminated and dark conditions).

We note that the background current in Fig. 2a has been elevated; it is a combination of channel doping and varying channel mobility near the Dirac point. Away from the Dirac point, say at $V_{gs}=0$ V where the conductivity is almost solely controlled by the charge density and less by the nonlinear channel mobility this is translated to a charge increase of $(7\times10^{-9}\text{A})/(1.6\times10^{-19}\text{A}) \rightarrow 4.4\times10^{10}/\text{cm}^2$ (since our sample area was 1 cm$^2$). This is approximately the number of carriers induced by 1 V of the gate (the so called geometrical effect) assuming an oxide thickness of 20 nm and alumina thickness of 50 nm. At the same time, the number of photons from a 30 mW/cm$^2$ laser at 532 nm is $10^{17}$ photons/s·cm$^2$. Another word, the graphene and the QDs each absorbed ~$10^{15}$ photons/s·cm$^2$. This is the number of excited carriers and at least for graphene, is larger than the saturation density at ~$10^{13}$ cm$^{-2}$ [29] and not that far from the saturation of SWCNT [30]. At large white light intensities, the photo-current decreased and became saturated (see SI section). All of these suggest that the channel photo-conductance and its carrier concentration <u>is not</u> solely dependent on the behavior of the QD but also on the photo-conductivity of the graphene itself.

*Coupling to surface modes:* The electromagnetic surface modes were bound on one side by the low index of perforated alumina/SiO$_2$ layer ($n_{Al2O3/SiO2}$~2) at the sample's bottom, and a 250 nm polymer/air layer from the sample's top ($n_{air}$~1.15). An approximation for the refractive index of graphene may be taken as, $n_{graphene}$~2.6-1.3i [24]). The electromagnetic radiation may be efficiently coupled with a surface mode when the wavevector of either the incident, or the scattered (or both) waves are at resonance with the wavevector of the perforated substrate [16]. Since the array pitch is smaller than the wavelength, a surface mode may become a standing wave, as well. The positions of the QDs are in-phase with the standing electromagnetic surface modes, resulting in an enhanced luminescence effect (Fig. 3).

The optimal launching conditions for a surface mode (or its interrogation) is achieved by a small tilt and in-plane rotation of the perforated substrate with respect to the p-polarized incident beam (The incident beam was polarized such that it had a polarization component perpendicularly to the sample's surface, or, consequently within the plane of incidence). Note that the array pitch is much smaller than the propagating wavelength and a bound surface mode is utilizing every other or even larger number of hole-planes. The tilt angle θ may be computed similarly to [24] as,

$$sin(\theta) = \frac{\lambda_0}{a}\sqrt{(\frac{4}{3})(q_1^2 - q_1 q_2 + q_2^2)} - n_{eff} \qquad (1)$$

Here, $\lambda_0$, is the incident or emitted wavelength, $a$, is the pitch for the holes array ($a$~100 nm), $q_1$ and $q_2$ are sub-integers (e.g., 1/3) representing the ratio between the array pitch and the propagating wavelength.

First we note that the equation *cannot be fulfilled* for the pump wavelength of 532 nm and $n_{eff}$~2.2 (which takes into account the refractive index of the graphene on the perforated alumina). Therefore, the peaks in Fig. 3 may only be attributed to the resonance effect at emission wavelengths. Upon tilting the sample, there are two symmetric FL peaks as per (1) and their mid-point is the true zero point. The FL peak(s) in Fig. 3a can be attributed to $q_1$=1/3, $q_2$=0 whereas, the peak(s) for Fig 3b may be attributed to $q_1$=1/3, $q_2$=1/4. Further proof of resonance condition is given in the SI section: the linewidth of the emitted radiation is seen to be clearly broaden and shifted at resonance conditions.

As a reference experiment, we measured QDs on a flat glass slide (not shown). Unlike Fig. 3, the FL signal decreased monotonically as a function of the tilt angle: while the flat substrate is tilted, the illuminated area is increased and the intensity per area is decreased as $cos(\theta)$, leading to reduction in the overall FL signal.

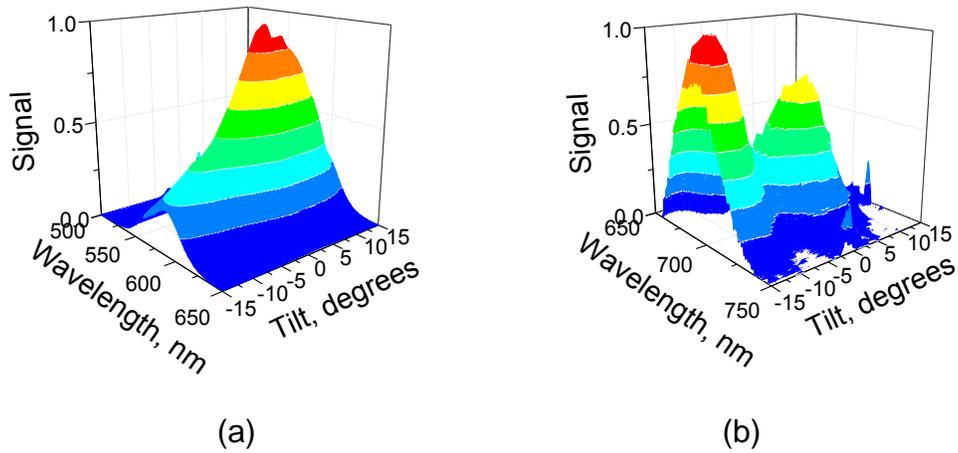

(a)                               (b)

Fig. 3. Fluorescence as a function of tilt angle. (a) For 590-nm QD. (b) For 670-nm QD. Note that the zero may be established at mid-point between the two symmetrical peaks. Thus the peak tilt angle is situated at $\theta \sim \pm 2^o$ for (a) and at $\theta \sim \pm 10^o$ for (b).

Fig. 4 shows the electrical and fluorescence (FL) data as a function of $V_{ds}$ at $V_{gs}=0$ V (Fig. 4a) and as a function of $V_{gs}$ at $V_{ds}=1$ V (Fig. 4b). The fluorescence monotonously decreased as a function of increasing $V_{ds}$. It also decreased as a function of increasing $V_{gs}$ for this limited range of $V_{gs}$. Similar results were obtained for a thicker substrate as shown in the SI section. Most puzzling is the effect induced by $V_{ds}$. As we show in the SI section, when the sample is at resonance with the optical surface mode then there seems to be no dependence of the fluorescence on either $V_{ds}$ or $V_{gs}$. In the following, we attempt to explain these results by the effect of the various capacitors involved in the process.

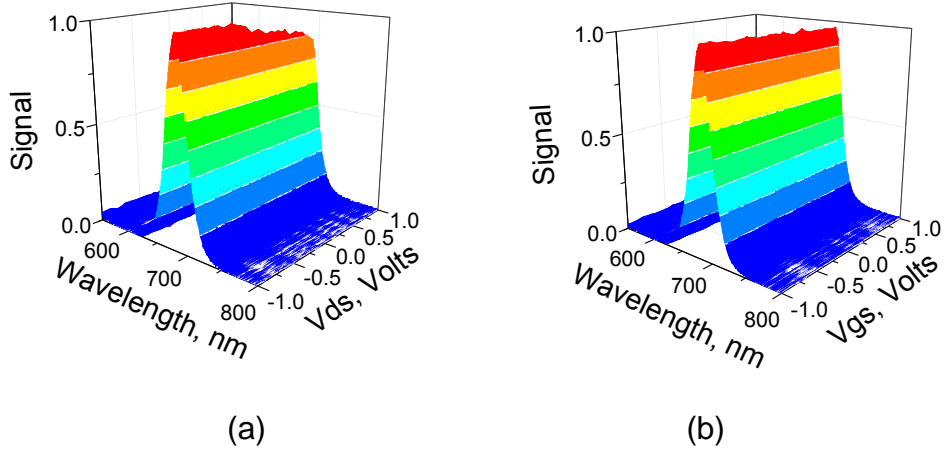

(a)                          (b)

Fig. 4. (a) FL as a function of $V_{ds}$ at $V_{gs}$=0 V and normal incident angle, $\theta$=0°. (b) FL as a function of $V_{gs}$ at $V_{ds}$=1 V and normal incident angle, $\theta$=0°. The oxide was 20 nm thick (hence the relatively lower values for $V_{gs}$.

**Dependence on $V_{gs}$:** From Fig. 3 it is clear that the graphene channel became even more n-type under uniform laser illumination at low laser intensity; the Dirac point has further shifted towards the negative $V_{gs}$ values. This could suggest that actual transfer of carriers from the QD 'doped' the graphene. We note though that a positive gate bias polarizes the excited electrons in the QD away from the surface and hence the probability of electron tunneling (as opposed to hole tunneling) is substantially reduced. If at all, the graphene would have been p-doped [6].

When considering a dipole, such as the QD near the graphene channel, the key parameter is the ratio between the QD's diameter, $d_0$, to the distance of its surface from the graphene channel, $d_B$. For example, assume that the graphene behaves as an infinite metallic-like surface versus a small dot of radius $d_0$. Also assume that the dot can be replaced by a dipole of size $d_0$. The attraction energy between the dot's charge, $Z_{QD}$ and its fictitious image is [30], $U_I \approx -Z_{QD} Z_{QD} \cdot d_0 / 4 d_B (d_B + d_0)$. If the dot diameter is much larger than $d_B$ then the attraction energy behaves as $1/d_B$. Similarly to conditional artificial dielectrics [31], these excited QD dipoles increase the gate-to-graphene capacitance and the result is a further 'doping' of the graphene channel. The induced charges are not diffused throughout the entire graphene channel but are more localized within a Thomas-Fermi distance of $Z_{QD} \cdot d_B$ [17]. This means that the charge density witin this localized area is rather large

**Dependence on Vds:** The surface potential of the graphene varies linearly along the channel. Specifically, $V_{ds}(x) \sim (x/L) \cdot (V_d - V_s)$; here $L$ is the channel length, $x$ is any point along the channel in this quasi 1D model (See the SI section for the equivalent circuit model). If the graphene is treated as a single resistive layer then a positive local surface bias counters the effect of a positive gate bias on the QDs and effectively de-polarizes QDs - this is not what we observe. If on the other hand, we treat the graphene as a capacitor with a self-capacitance of µF/cm$^2$, then positive $V_{ds}$ injects positive charges to one layer of this capacitor and the other layer interfacing the gate further the impact of the gate bias. Thus, positive $V_{gs}$ and $V_{ds}$ values negatively dope the graphene and suppress the fluorescence via screening. Finally, we observed that when the emission radiation is at resonance with the porous substrate its signal was unaffected by the electrical bias (see SI section). FRET is crucially depending on the life-time of the donors (QD in our case) which ought to be longer than the acceptor channel (the excited e-h pairs in the graphene). Specifically, $k_{FRET} = 1/\tau_{DA} - 1/\tau_D$, where $\tau_{DA}$ and $\tau_D$ are the life-time of the donor-acceptor route and the fluorescence of a stand-alone donor's route, respectively. Increasing the emission rate for the QD, at resonance with the substrate hinders the energy transfer of energy to the graphene and increased fluorescence. Actual carrier transfer would not be affected by such resonance considerations at the emission frequencies.

The effect of leakage current has been assessed; it has been found that $I_{ds}$ as a function of $V_{gs}$ in the range of [-5,5] V at $V_{ds}$=0 V was at least a factor of 10 smaller (or on the order of 0.1 nA) than the current level at $V_{ds}$=0.01 V (which was on the order of nA).

In summary, by using graphene as an optical and electrical surface guide in an FET construction, and by coupling the graphene channels with commensurate, yet individual quantum semiconductor dot array, we have demonstrated a unique electro-photonic structure which may find applications in communication and sensing systems.

**Acknowledgement**

**Graphene channels interfaced with an array of individual quantum dots**

Xin Miao and Haim Grebel*

*Electronic Imaging Center and ECE Dept., New Jersey Institute of technology (NJIT), Newark, NJ 07102, USA. grebel@njit.edu*

Supplementary Information

**Photo-current:** The $I_{ds}$-$V_{gs}$ curve for QDs interfaced graphene channels under uniform white light illumination and under dark conditions is shown in Fig. S1. Typically, a minimum in the $I_{ds}$-$V_{gs}$ curve (Fig. 2b) signifies the condition where the Fermi level of the graphene is situated at the Dirac point (the conduction band is empty while the valence band is full, at zero temperature). Here, the QDs are partially excited at room temperature and the resulting gate effect makes the graphene channel more of an n-type at $V_{gs}$=0 V. Illumination by a laser, or white light resulted in a Dirac point shifting towards the negative $V_{gs}$ values (namely, the channel becomes even more of n-type at $V_{gs}$=0).

Fig. S1 shows the effect of a channel under white light illumination at intensities of 380 and 440 mW/cm$^2$, respectively. The white-light beam illuminated the entire sample area. Shown is the current difference (current under white-light minus the current at dark conditions) as a function of $V_{gs}$ and $V_{ds}$. The channel here was deposited on a 150 nm thick oxide, hence the relatively large $V_{gs}$ values.

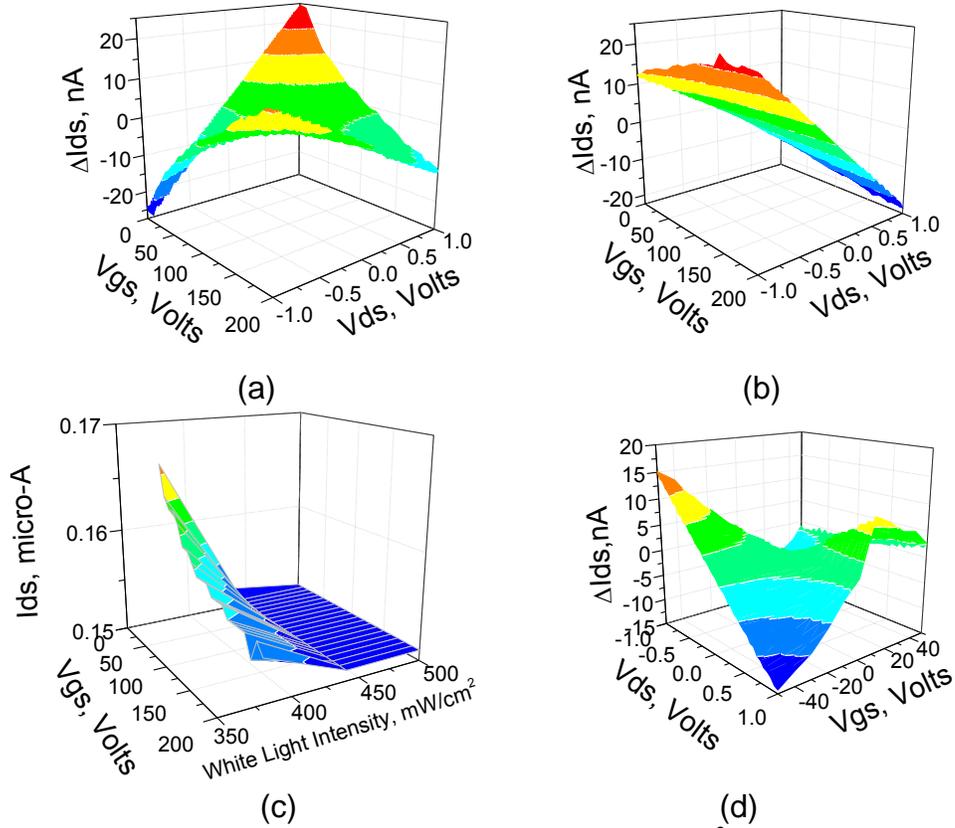

Fig. S1. (a) Channel conductivity under white light (380 mW/cm$^2$) and dark conditions. Plotted is the difference in channel current as a function of $V_{gs}$ and $V_{ds}$. The thicker oxide of 150 nm resulted in larger $V_{gs}$ values. Compared with $V_{gs}$=0 V, $V_{gs}$=200 V the differential current changes direction. (b) The differential current at larger intensity of white light illumination (440 mW/cm$^2$); the differential current has a negative trend for both $V_{gs}$=0 and $V_{gs}$=200 V. (c) Channel current, $I_{ds}$ as a function of $V_{gs}$ and white light intensity at $V_{ds}$=0.5V. (d) Another sample illuminated by white light: the peak in the differential current as a function of $V_{gs}$ can be explained by the position of the Dirac point, close to $V_{gs}$=+20 V.

As can be seen from the S1(a,b), one can identify two trends: (1) when the light intensity is relatively small. The $\Delta I_{ds}$-$V_{ds}$ curve slop is constant and positive for small values of $V_{gs}$ while negative for large values of $V_{gs}$. This suggests that the photocurrent properties (which is proportional to the differential current) may have more to do with the graphene characteristics rather than with the QDs' because, as the 'doping' of the graphene becomes larger and the conductive states become occupied, the photo-assisted transition of electrons require larger optical energies (which are limited by the spectral range of the white light source). Had the doping of graphene originated from ionized QDs (similar to doped semiconductors), then Fig. S1c would exhibit a reverse behaviour: as the white light intensity is increased the photo-current would increase as well, until being saturated. (2) As we increase the white-light intensity, the overall trend of the $\Delta I_{ds}$-$V_{ds}$ curve became negative for all $V_{gs}$ values. This again suggests a saturation of the photo-assisted transitions in the graphene. Indeed, Fig. S1c exhibits current saturation beyond intensity values of 440 mW/cm$^2$. We note that the FL experiments were conducted with a focused laser beams whose intensity was on the order of 10$^5$ W/cm$^2$. Finally, we show another sample where the Dirac point was situated at $V_{gs}$>0 which could be attributed to an unusual surface potential as a result of sample preparation.

**Fluorescence Measurements:** Enhanced peak luminescence at resonance and at off-resonance conditions are shown in Fig. S2. Shown are the spectral curves at tilt angle of θ=−5º (close to emission minima) and at tilt angle of θ=−13º (close to the emission maxima). Clearly seen is a line broadening of more than 25% which is attributed to a Purcell's effect (namely, an increase in the density of states when the luminescing wavelength is at resonance with a cavity). The Purcell's effect alludes to a decrease of the life-time of the excited e-h pair in the QD as measured for a similar system of QD in AAO [S1]. There is also a small but clear peak shift due to the particular hole-array pitch involved in the resonance condition.

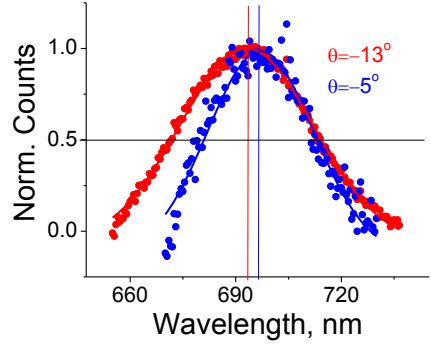

Fig. S2. Comparison of normalized spectral curves at tilt angle of θ=−5º (close to emission minima) and at θ=−13º (close to the emission maxima) clearly exhibiting line broadening of FWHM from ca 32.3 to 41.5 nm. The solid lines are Gaussian fits to the curves. The peak shift was −4±0.3 nm.

Suppression of the fluorescence [S2] as a function of $V_{gs}$ may be attributed to the change in the channel conductivity [S3]; as the channel became more conductive, the fluorescence quenches. This is a local effect due to the conductivity screening by the graphene at the vicinity of localized QD.

The change in FL as a function of $V_{ds}$ may be understood with the model shown in Fig. S3. The surface potential, $V_{ds}(x)$ on the graphene channel varies linearly from source (typically at ground) to drain (held at a potential $V_d$). Since the waveguide is all but surface, one may consider the surface potential of the guide as a function of $V_{ds}$, as $V_{ds}(x)=(V_d−V_s)(x/L)$ where $L$ is the channel length. In general one may identify three effects: (1) the effect of the gate capacitor, $C_g$; the channel is considered as 'electrically doped'. (2) The effect of the excited QD on the local gate bias via its own capacitor $C_{QD}$; the channel is further 'electrically doped' since it is polarized by the gate in a direction opposite to the external gate field. (3) The effect of the local surface potential of the graphene guide itself. Photo-excitation has two effects: (1) excitation of a dipole within the QD. The dipole is polarized by the gate bias similarly to artificial dielectrics and the overall effect is to increase $C_g$ and hence the 'doping' of the graphene channel (see Fig. 2 in the main text). (2) Excitation of electrons within the graphene. The QDs have a shell barrier (ZnS) and are coated with a polymer (octadecylamine) to prevent agglomeration while in suspension and therefore, a direct contact between the QD and the graphene is less likely. In cases where the dot is in close proximity to the channel, then the probability of electron tunneling from the QD to the graphene may be written as, $\sim exp[-2d'_B(\Phi_b - eV)^{1/2}]$ where $d'_B$ is the equivalent barrier width between graphene and QD, $\Phi_b$ is the barrier height, $V = V_{ds}(x) - V_g$ (the negative sign for $V_g$ is due to the gate effect at the graphene surface) and $e$ is the electronic charge. Overall, tunneling negates the effect of charge polarization at the QD which is contrary to our experiments.

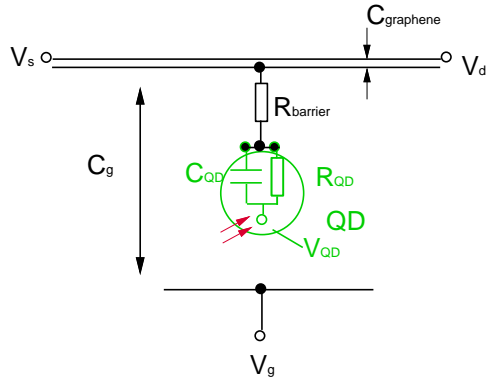

Fig. S3. A circuit model that illustrates the various effects on the graphene channel. The source, $V_s$ is typically grounded. $C_g$ is the capacitor between the gate and the graphene channel; as the gate bias becomes more positive, the graphene guide becomes more negative (or more n-doped). $C_{QD}$ is the equivalent dot capacitor (whose polarization negates that of the $C_g$) and $R_{QD}$ is the equivalent dot resistor (which is quite large). $R_{barrier}$ is the resistance between the dot and the graphene channel.

The effect of bias on the thicker oxide is shown in Fig. S4.

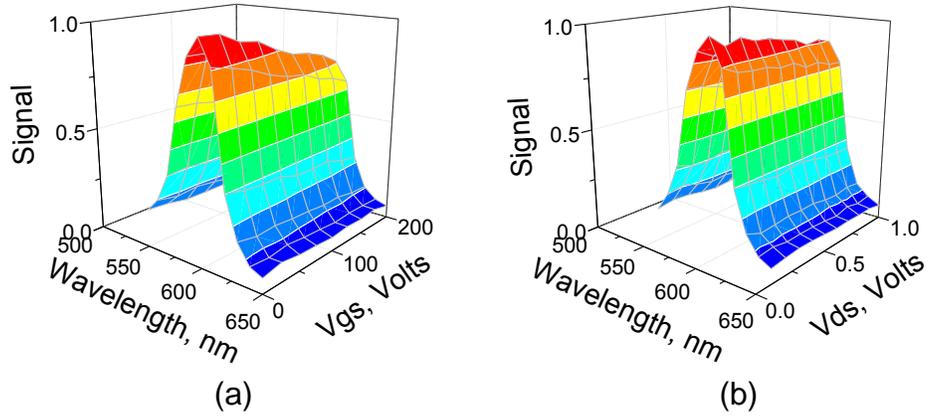

(a)                          (b)

Fig. S4. The effect of bias on QD interfaced graphene on a 150 nm thick oxide.

Finally, we studied the FL as a function of $V_{gs}$ at two tilt angles (namely, at on- and off-resonance with respect to the hole-array). At off-resonance, the FL exhibited a monotonous decline of overall 5% as a function of $V_{gs}$, whereas it was flat at resonance conditions. Similar trend was found for FL vs $V_{ds}$ at $V_{gs}=-5$ (close to the negative photo conductance identified in Fig. 2a,b).

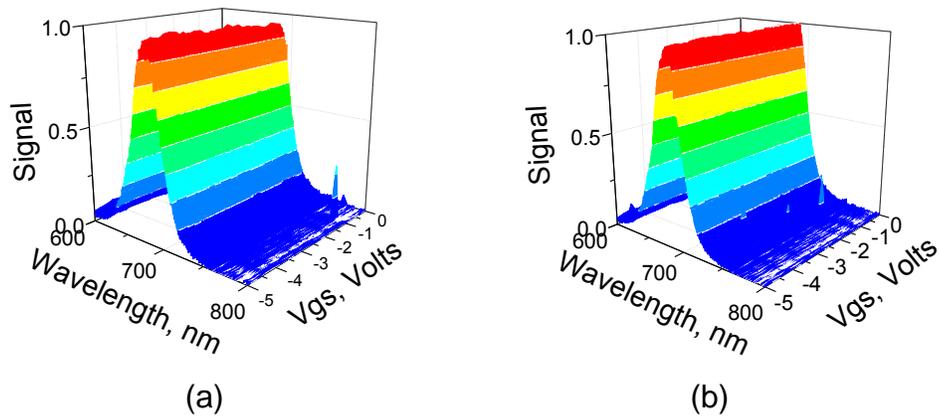

(a)                          (b)

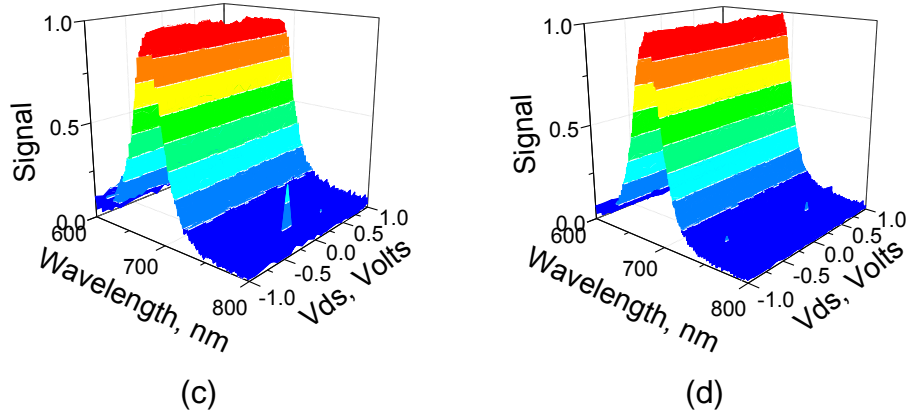

(c)                          (d)

Fig. S5. (a,b) FL as a function of $V_{gs}$ ($V_{ds}$=0.3 V) and (c,d) as a function of $V_{ds}$ ($V_{gs}$=−5 V), at off-resonance (tilt at $\theta$=0°) and at on-resonance (tilt at $\theta$=−15°), respectively. In (a,c), the FL change between minima and maxima is 5%±0.7%. In (c), there are two small symmetric peaks at $V_{ds}$=±0.5.